\documentclass{article}[12pt]

\usepackage{multirow}
\usepackage{hyperref}
\usepackage{graphicx}
\usepackage{color}
\usepackage{enumitem}
\usepackage{mathrsfs}
\usepackage[T1]{fontenc}
\usepackage[ruled,vlined]{algorithm2e}  
\usepackage{amssymb,amsmath,amsfonts,amsxtra,amsthm}
\usepackage[a4paper, total={6in, 8in}]{geometry}

\makeatother
\begin{document}

\title{The robusTest package: two-sample tests revisited}
\author{Sinda Ammous$^{(1)}$, Olivier Bouaziz$^{(1, 2)}$, Anatole Dedecker$^{(3)}$,\\ J\'er\^ome Dedecker$^{(1, 2)}$, Jonathan El Methni$^{(1, 2)}$, Mohamed Mellouk$^{(1, 2)}$ and Florence Muri$^{(2)}$}
\date{\small $^{(1)}$ Universit\'e Paris Cit\'e, CNRS, MAP5, UMR 8145, 75006 Paris, France \\ 
$^{(2)}$   IUT de Paris -- Rives de Seine, 75016 Paris, France \\ 
$^{(3)}$ Université Paris-Saclay, Paris, France}

\maketitle

\abstract{
The R package {\it robusTest} offers corrected versions of several common tests in bivariate statistics. 
We point out the limitations of these tests in their classical versions, some of which are well known such as robustness or calibration problems, and provide simple alternatives that can be easily used instead. The classical tests and their robust alternatives are compared through a small simulation study. 
The latter emphasizes the superiority of robust versions of the test of interest.
Finally, an illustration of correlation's tests on a real data set is also provided.
}

\section{Introduction}

\indent
In this article, we consider several usual tests in bivariate statistics, which are taught
in many scientific courses at various levels.\\
\indent
Our first goal is to illustrate, with the help of mathematical considerations and simulations, that many
of these tests are either not very robust (i.e. they do not work outside the very strict framework
in which they have been defined), or badly calibrated 
(i.e. we can find simple examples for which the null hypothesis is true, but the type I error rate is not the one announced).\\
\indent
Of course, we are not the only ones or the first ones to have noticed this, and in some cases
valid solutions have been proposed (see for example the famous article by Welch \cite{We} 
about the test of equality of two expectations in the case where the variances are unequal, the Welch test being (asymptotically) robust to non-normality).\\
\indent
Nevertheless, we felt it was important to return to these issues for at least two reasons:\\
\indent
- First,  these tests are not only mathematical objects on which students can discover the basic principles of statistical analysis, they are also very often used in practice (for example in biomedical research articles). 
It is therefore important to come back to the limits of these tests, which are not always well indicated (especially in the case of so-called "non-parametric" tests).\\
\indent
- Secondly, because it is often very easy to modify these tests in order to make them more robust or
(asymptotically) well calibrated. 
The modification presented is each time based on the calculation of the limiting variance of the test statistic under the null hypothesis. Renormalizing by an estimator of the
standard deviation, we obtain a robust or asymptotically well-calibrated version thanks to the
central limit theorem. Of course, the justification is asymptotic, but we will illustrate that, on 
simulated examples, it is always interesting to correct these tests, even for relatively small sample sizes.
It seems to us that these modified tests, which are easy to describe and to implement, should be 
systematically pointed out to the students, and also to researchers from other disciplines unfamiliar with these issues.\\
\indent
Our second objective is to present the {\it robusTest} package, which implements robust modifications of the usual bivariate statistics tests. As we will see, the functions of the {\it robusTest} package, as well as their syntax, are very close to the functions of the {\it stats} package, so that regular users
of the R software can use and compare them easily. This package is already used at  Université Paris Cité and  Université d'Evry Val d'Essonne to illustrate second or third year postgraduate courses on bivariate statistics. \\
\indent
The article is organised as follows: in  Section 2, we present the robust versions of the Pearson (see~\cite{Fi}), Kendall (see~\cite{K}) and Spearman (see~\cite{HP}) correlation tests. For completeness, we also provide in the same section the Kolmogorov-Smirnov type test of independence for continuous variables (see~\cite{BKR}). In Section 3, we describe the test of equality of conditional variances of $X$ knowing $Y$, when $Y$ is a categorical variable with several levels. This procedure is based on the James-Welch ANOVA which we briefly recall (see~\cite{J} and  \cite{We2}).  In Section 4, we present the robust version of the two sample Mann-Whitney test for stochastic dominance (see~\cite{MW}). In Section 5, we describes several ways to test the stochastic dominance in the case of paired two samples (see~\cite{vdv}). At the end of each section or subsection, we provide the R-functions of the {\it robusTest} package for each corrected test. Finally, in the last section, we illustrate on a real data set the functions of the {\it robusTest} package to test the correlation, and we compare the outputs to those of the usual tests. 

\section{Robust tests for testing correlation and independence}
\label{SeRTCI}

In the following four subsections we respectively focus on the correlation tests of Pearson, Kendall and Spearman, and on the independence test of Kolmogorov-Smirnov type.

\subsection{Pearson's correlation test}
\label{Sec1}

Let $(X_1,Y_1), \ldots, (X_n,Y_n)$ 
be independent and identically distributed (i.i.d) copies of the random pair $(X,Y)$, where $X,Y$ are real-valued random variables.
We suppose that $X_i$ and $Y_i$ have finite moments of order 2 and we denote ${\mathbb E}(X_i)=\mu_X$ and ${\mathbb E}(Y_i)=\mu_Y$.
Our aim is to test 
$$
\text{the null hypothesis} \quad  H_0: \rho= 0 \quad \text{against the alternative hypothesis} \quad H_1: \rho \neq 0
$$
where $\rho$  is the correlation coefficient between 
$X_i$ and $Y_i$.
To this end we often use
the Pearson test statistic  (see for instance Fisher \cite{Fi}):
$$
T_n=  \frac{ \hat \rho_n}{\sqrt{\frac{1- \hat \rho_n^2}{n-2}}}
\quad \mbox{where} \quad 
\hat \rho_n= \frac{\sum_{k=1}^n (X_k-\bar X_n)(Y_k-\bar Y_n)}{\sqrt{\sum_{k=1}^n (Y_k - \bar Y_n)^2} \ \sqrt{ \sum_{k=1}^n (X_k -\bar X_n)^2}} \,
$$
is the empirical correlation coefficient and where $\bar X_n$, $\bar Y_n$ are the empirical means.\\
\indent
The interest of this statistic being that, in the case where the couple $(X_i, Y_i)$ is Gaussian, its exact distribution  under $H_0$ is known: it is the Student $St(n-2)$-distribution.
If the pair $(X_i, Y_i)$ is not Gaussian, we can also easily show that, if $X_i$ is independent of $Y_i$ (and if $X_i$ and $Y_i$ are not constants), then $T_n$ converges in distribution to the standard normal distribution ${\mathcal N}(0,1)$.  
Therefore, if $X_i$ and $Y_i$ are independent,  the rejection region $R_{n, \alpha}= \{ |T_n|> c_\alpha \}$ where $c_{\alpha}$ is the quantile of order $1-(\alpha/2)$ of the  ${\mathcal N}(0,1)$ distribution, is such that its probability tends to 
 $\alpha$ as $n \rightarrow \infty$. But this is not the case in general under $H_0$, which of course does not imply
independence of the variables.\\
\indent
We can therefore make the following conclusion:
 in a general context, the Pearson test is not well calibrated to test the null hypothesis $\rho=0$. See also
\cite{EN} for a similar observation. Outside the strict framework of the linear model (Gaussian or not), it is therefore preferable to use the intuitive statistic
$$
  T'_n= \frac{\sum_{k=1}^n (X_k-\bar X_n)(Y_k-\bar Y_n)}{\sqrt{\sum_{k=1}^n (Z_k-\bar Z_n)^2}}
  \quad \mbox{where} \quad  Z_i=(X_i-\bar X_n)(Y_i-\bar Y_n).
$$
Under the assumption $ 0 < {\mathbb E} \left ( (X_i - \mu_X)^2 (Y_i- \mu_Y)^2 \right ) < \infty$, a direct application of the central limit theorem and
Slutsky's lemma provides the convergence in distribution  of $T'_n$ under $H_0$ to the standard normal distribution. The rejection region of $H_0$ is 
$R'_{n, \alpha}= \{ |T'_n|> c_\alpha \}$ where $c_{\alpha}$ is the quantile of order $1-(\alpha/2)$ of the  ${\mathcal N}(0,1)$ distribution, which provides a test
asymptotically well calibrated.\\

\noindent{\bf Remark.}
It is easy to see that, in the Gaussian case and under $H_0$, the distribution of the statistic $T'_n$ does not depend on the expectation and variance parameters, and can therefore be tabulated. In practice, in the {\it{robusTest}} package, we use this table of quantiles for $n<130$ and the quantiles of the Student $St(n-2)$  distribution for $n\geq 130$ (simulations show that the quantiles of the Student $St(n-2)$  distribution are close to those of the distribution of the statistic $T'_n$ under $H_0$ in the Gaussian case).\\ 

\noindent
The corresponding R function of the {\it{robusTest}} package is: \texttt{cortest(,method="pearson")}.

\medskip

\noindent{\bf Remark.} 
Starting from the quantity 
$$
\frac{\sum_{k=1}^n (X_k-\bar X_n)(Y_k-\bar Y_n)-n\text{Cov}(X,Y)}{\sqrt{\sum_{k=1}^n (Z_k-\bar Z_n)^2}}\, ,
$$ which converges to the ${\mathcal N}(0,1)$ distribution as $n\rightarrow \infty$, we can easily obtain a confidence interval for  $\text{Cov}(X,Y)$, the covariance between $X$ and $Y$. As an output of the command \texttt{cortest(,method="pearson")}, we  propose instead a confidence interval for Pearson's $\rho$ coefficient. This confidence interval is based on the central limit theorem for the empirical estimator of  $(\text{Cov}(X,Y), \text{Var}(X), \text{Var}(Y))$ and the delta method applied to the function $h(x,y,z)=x/\sqrt{yz}$ from ${\mathbb R}^3$ to ${\mathbb R}$.  There may be a difference between the result based on the robust Pearson test and whether or not 0 is in the confidence interval of $\rho$. In this case, the test should be preferred since the confidence interval based on the central limit theorem + the delta method is a priori less precise.

\subsection{Kendall's correlation test}
\label{Sec2}

\setcounter{equation}{0}

The context is the same as in the previous paragraph
$(X_1,Y_1), \ldots, (X_n,Y_n)$ are i.i.d.copies of the pair $(X,Y)$,  where $X,Y$ are real-valued random variables. We assume moreover that the variables are continuous. We want to know if $X_i$ and $Y_i$ tend to vary in the same direction or in the opposite direction. Let then 
 $$
   \tau= 2\left({\mathbb P}((X_2-X_1)(Y_2-Y_1) >0) -0.5\right)
 $$
 be Kendall's  correlation coefficient \cite{K}. As for Pearson's correlation coefficient, $\tau$ is between $-1$ and 1. $X_i$ and $Y_i$ are positively correlated in the sense of Kendall when $\tau > 0$ and negatively if $\tau <0$.
 To test
 $$
  H_0: \tau= 0 \quad \text{against} \quad H_1: \tau \neq 0 
$$
Kendall  \cite{K} therefore proposed to count the number of concordant pairs
  (i.e. for which the product $(X_i-X_j)(Y_i-Y_j)$ is strictly positive), which leads to the statistic
 $$
 T_n=  \frac{1}{n(n-1)} \sum_{i=1}^n \sum_{j=1, j \neq i}^n \left ( {\mathbf 1}_{(X_i-X_j)(Y_i-Y_j)>0}  - 0.5\right )  \, .
 $$
 It is quite easy to see that, if $X_i$ and $Y_i$ are independent, then the distribution of $T_n$ is distribution-free (i.e. does not depend on the distribution of  ($X_i$,$Y_i$)), by reducing to two independent sequences $U_1, \ldots, U_n$ and $V_1, \ldots, V_n$ of i.i.d. random variables with standard uniform distribution ${\mathcal U}([0,1])$. Consequently, if $X_i$ and $Y_i$ are independent, $T_n$ is distributed according to a  known and tabulated distribution.
 Kendall's test is constructed from the quantiles of this distribution. However if $X_i$ and $Y_i$ are not independent, the distribution of the statistic $T_n$ has no reason to be distribution free under $H_0$
 (it depends on the joint distribution of $(X_i,Y_i)$). We can therefore make the following conclusion:
in a general context, Kendall's test is not well calibrated to test $\tau=0$.\\
\indent
 We can nevertheless solve (asymptotically) this problem, by considering the limiting distribution of
 $\sqrt n T_n$ under $H_0$. Starting from the Hoeffding decomposition of the $U$-statistic $T_n$ (see \cite{H} or \cite{vdv}, example 12.5), we see that, under $H_0$,
$\sqrt n T_n$ converges in distribution to the ${\mathcal N}(0, V)$ distribution, with
$$
V=4 {\mathrm{Var}(F(X_i,Y_i) + H(X_i,Y_i))} \, ,
$$
where $F(x,y)= {\mathbb P}(X_i <x, Y_i <y)$ and $H(x,y)={\mathbb P}(X_i >x, Y_i >y)$. The empirical estimator of
$V$ is then
$$
 V_n=  \frac{4}{n-1} \sum_{k=1}^n \left (F_n(X_k, Y_k) + H_n(X_k, Y_k) - \bar{F}_n-\bar{H}_n \right )^2
$$
where
\begin{align*}
F_n(x,y) & = \frac 1 n \sum_{k=1}^n  {\mathbf 1}_{X_k < x, Y_k <y}\, , \quad   H_n(x, y) =\frac 1 n \sum_{k=1}^n  {\mathbf 1}_{X_k > x, Y_k >y} \\
\bar F_n & = \frac 1 n \sum_{k=1}^n
F_n(X_k, Y_k)  \, , \quad  \bar H_n  =\frac 1 n \sum_{k=1}^n  H_n(X_k, Y_k)\, .
\end{align*}
Finally, under $H_0$, 
$$ K_n=\frac{\sqrt n T_n}{ \sqrt{V_n}} \  \ \text{converges in distribution to the ${\mathcal N}(0, 1)$ distribution.}$$
The rejection region of the corrected Kendall test is then
$R_{n, \alpha}= \{ |K_n|> c_\alpha \}$ where $c_{\alpha}$ is  the quantile $1-(\alpha/2)$ of the ${\mathcal N}(0,1)$ distribution, which provides a test asymptotically well calibrated (see \cite{Am} for more details).\\

\noindent
The corresponding R function  of the {\it{robusTest}} package is:
\texttt{cortest(,method="kendall")}.

\subsection{Spearman's correlation test} In the same context as for the Kendal test statistic of the previous paragraph, 
we can also correct Spearman's correlation test (as described by Hotelling and Pabst \cite{HP}), which tests
 $$
  H_0: \rho_S= 0 \quad \text{against} \quad H_1: \rho_S  \neq 0 
$$
where $\rho_S$ is the correlation coefficient between the variables $F_X(X_i)$ and $F_Y(Y_i)$ uniformly distributed over
$[0,1]$ (here $F_X(x)= {\mathbb P}(X \leq x)$ and $F_Y(y)= {\mathbb P}(Y\leq y)$).
Like the Kendall and Pearson tests, this test is not well calibrated
if the variables $X_i$ and $Y_i$ are not independent.\\
\indent
As with Kendall's test,  Hoeffding \cite{H} showed that Spearman's test statistic can be expressed using a
$U$-statistic, from which he deduced à central limit theorem for the normalized statistic, with an exact expression of the limiting variance. As in the previous paragraph we can estimate this limit variance (by taking the empirical estimator), and then obtain a corrected test which is asymptotically well calibrated. The expression of the limit variance being  more complicated than that of Kendall's statistic, we do not give all the details  here.\\
 
\noindent
The corresponding R function of the {\it{robusTest}} package is: \texttt{cortest(,method="spearman")}.

\subsection{Kolmogorov--Smirnov statistic for testing independence}
The context is the same as in the previous subsection: 
  $(X_1,Y_1), \ldots, (X_n,Y_n)$ are independent copies of a couple $(X,Y)$ of real-valued random variables, and we assume that
  the variables $X$ and $Y$ are continuous. 
We want to test 
 $$
  H_0: X ~\mbox{and}~ Y ~\mbox{are independent}~ \quad \text{against} \quad H_1: X ~\mbox{and}~ Y ~\mbox{are not independent}~
$$
To answer this question, one can use the Kolmogorov-Smirnov statistic (as described for instance in Blum, Kiefer and Rosenblatt  \cite{BKR})
$$
KS_n= \sup_{s, t \in {\mathbb R}} \sqrt n \left |  \frac 1 n \sum_{i=1}^n  {\bf 1}_{X_i\leq t, Y_i\leq s} - F_{n,X}(t) F_{n,Y}(s)\right | \, ,
$$ 
where 
$$
F_{n,X}(t)=\frac 1 n \sum_{i=1}^n  {\bf 1}_{X_i\leq t}\, , \quad F_{n,Y}(s)=\frac 1 n \sum_{i=1}^n  {\bf 1}_{Y_i\leq s}\, .
$$
It is clear that, under the hypothesis $H_0$: $X_i$ is independent of $Y_i$, the statistic $KS_n$ is distribution-free (since the variables are continuous, one can go back to the case where $X_i$ and $Y_i$ are uniformly distributed over $[0,1]$). On another hand, one can easily check that $KS_n$ is asymptotically equivalent to
$$
KS'_n=\sup_{s, t \in {\mathbb R}} \sqrt n \left | \frac 1 n \sum_{i=1}^n ( {\bf 1}_{X_i\leq t}-F_X(t)) ({\bf 1}_{Y_i\leq s}-F_Y(s))\right |\, .
$$
Now, as proved in \cite{BKR}, $KS'_n$ (and hence $KS_n$) converges in distribution under $H_0$ to the supremum of a Gaussian process.
The rejection region of the Kolmogorov-Smirnov test of independence is then $R_{n, \alpha}= \{ KS_n> c_\alpha \}$ where $c_{\alpha}$ is the exact quantile of order $1-\alpha$ of the distribution of $KS_n$ under $H_0$ (note that, for small $n$, the quantity ${\mathbb P}_{H_0}(R_{n, \alpha})$ is in general not exactly equal to $\alpha$, because the distribution of $KS_n$ under $H_0$ is a discrete distribution). These quantiles (or the $p$-value of the test) can be easily estimated via a basic Monte-Carlo procedure.\\

\noindent
The corresponding R function of the {\it{robusTest}} package is:
\texttt{indeptest()}.

\medskip

\noindent {\bf Remark:} When the continuous variables $X_i$ and $Y_i$ are observed with too rough a rounding, the statistics of Kendall, Spearman or Kolmogorov-Smirnov can behave badly (because they involve quantities of the type ${\bf 1}_{X_i>Y_j}$). In the {\it{robusTest}} package, we add the possibility to correct this problem by a simple randomization procedure (if there is a tie, we toss heads to see if the indicator is 0 or 1); to do so, it suffices to use the argument \texttt{ties.break="random"}. This remark is also valid for the Mann-Whitney test which will be presented in one of the following sections.\\

\noindent
In order to highlight the differences between the classicals tests and the corrected tests, we will consider in the next section two simulation scenarios.

\subsection{Simulation study}

\medskip

\begin{figure}[htbp]
\centering
\includegraphics[width=11.5 cm, height=9 cm]{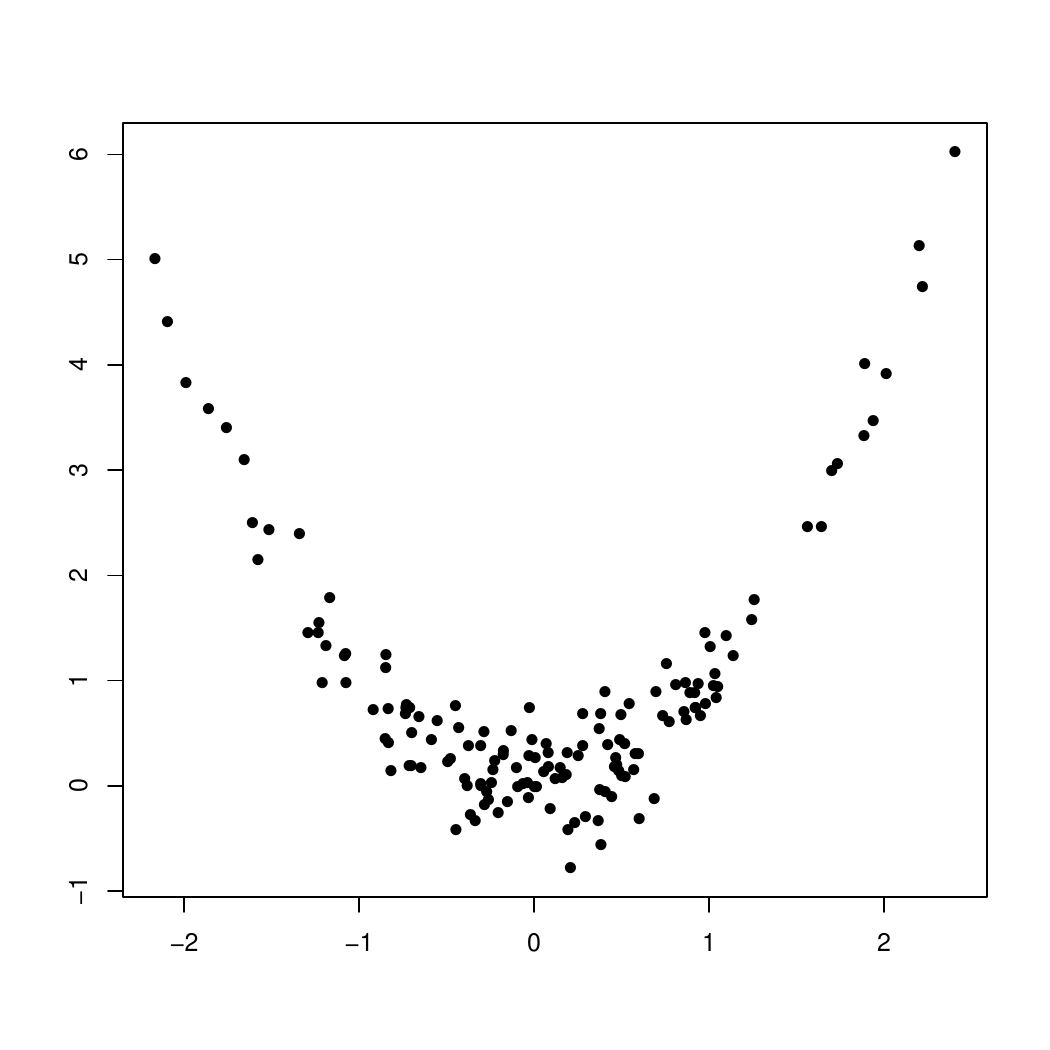}
\caption{Scatter plot of 150 couples $(x_i, y_i)$ drawn according to model \eqref{mod1}}
\label{fig:mod1}
\end{figure}

\noindent{\bf First scenario:} 
we simulate, for different values of $n$, i.i.d. pairs $(X_i, Y_i)_{1 \leq i \leq n}$ according to the model
\begin{equation}\label{mod1}
Y_i= X_i^2 + 0.3 \varepsilon_i
\end{equation}
 where the $(X_i)_{1 \leq i \leq n}$ and the $(\varepsilon_i)_{1 \leq i \leq n}$ are two independent sequences of i.i.d variables with  ${\mathcal N}(0,1)$ distribution (see Figure \ref{fig:mod1}).
One can easily see that, for this model, $\rho=\tau=\rho_S=0$. Hence $X_i$ and $Y_i$ are not correlated in the sense of Pearson, Kendall or Spearman; but of course, they are not independent.

\indent
We simulate $N=3000$ samples of size $n$ according to model \eqref{mod1}, and for each test we indicate the frequency of rejection of
$H_0$ at level $5\%$. We consider seven tests : the usual Pearson (usual P), Kendall (usual K) and Spearman (usual S), and the Robust Pearson (robust P), Kendall (robust K) and Spearman (robust S), to test $H_0: \rho=0$, $H_0: \tau=0$ and $H_0: \rho_S=0$ respectively; the Kolmogorv-Smirnov independence test (KS indep) to test $H_0: X$ and $Y$ are independent. The results are given in Table \ref{tab1} (level $\alpha= 5 \%$). 

 \begin{table}
\center
\begin{tabular}{|c|c|c|c|c|c|c|c|c|c|c|}
\hline
 $n $  &    30  & 40 & 50 & 60 & 70 & 80  & 90 & 100 & 200 & 300 \\
\hline 
usual P  & 0.361  &  0.376 &  0.365 & 0.369 & 0.38 & 0.353 & 0.366 &0.362 & 0.374 &0.384\\
\hline
robust P &0.055 &0.047 & 0.048 &0.052 & 0.055 & 0.05 &0.049&0.049 &0.051  & 0.054\\
\hline
usual K &0.174 &0.184 & 0.183 &0.201 & 0.201 & 0.202 &0.187&0.194 &0.203  & 0.194\\
\hline
robust K &0.077 &0.067 & 0.069 &0.066 & 0.064 & 0.057 &0.057&0.052 &0.052  & 0.051\\
\hline
usual  S &0.123 &0.124 & 0.132 &0.135 & 0.145 & 0.131 &0.131&0.126 &0.139  & 0.137 \\
\hline
robust S &0.113 &0.096 & 0.089 &0.082 & 0.078 & 0.071 &0.068&0.061 &0.057  & 0.056 \\
\hline
KS indep &0.732 &0.934 & 0.989 &0.999 & 1 & 1 & 1& 1 & 1  & 1 \\
\hline
\end{tabular}
\caption{Frequencies of rejection of the seven tests at level $5\%$ for model  \eqref{mod1}.}
\label{tab1}
\end{table}

From Table \ref{tab1}, 
we can notice that the three usual correlation tests are poorly calibrated, with rejection frequencies of $H_0$ around
$37\%$ for usual P, $19\%$ for usual K, and $13\%$ for usual S, instead of the expected $5\%$. 
We can also remark on this set of simulations, that it is always preferable to use robust tests, whose rejection frequencies are always lower than those of the usual tests,  and close to $5\%$ for $n\geq 100$. In fact, for robust P, the frequencies of rejection are always between $4.7\%$ and $5.5\%$ for $n\geq 30$; for robust K, the frequencies of rejection are always between $5\%$ and $7\%$ for $n\geq 40$; for robust S, the frequencies of rejection are always between $5\%$ and $7\%$ for $n\geq 90$. The fact that the rejection frequencies of robust S converge more slowly towards $5\%$ can perhaps be explained by the fact that the asymptotic variance term of the Spearman statistic is quite complicated, and is therefore more difficult to estimate. 
To conclude, unsurprisingly, the independence test detects very well the non-independence of $X$ and $Y$, systematically as soon as $n\geq 70$.

\medskip

\begin{figure}[htbp]
\centering
\includegraphics[width=11.5 cm, height=9 cm]{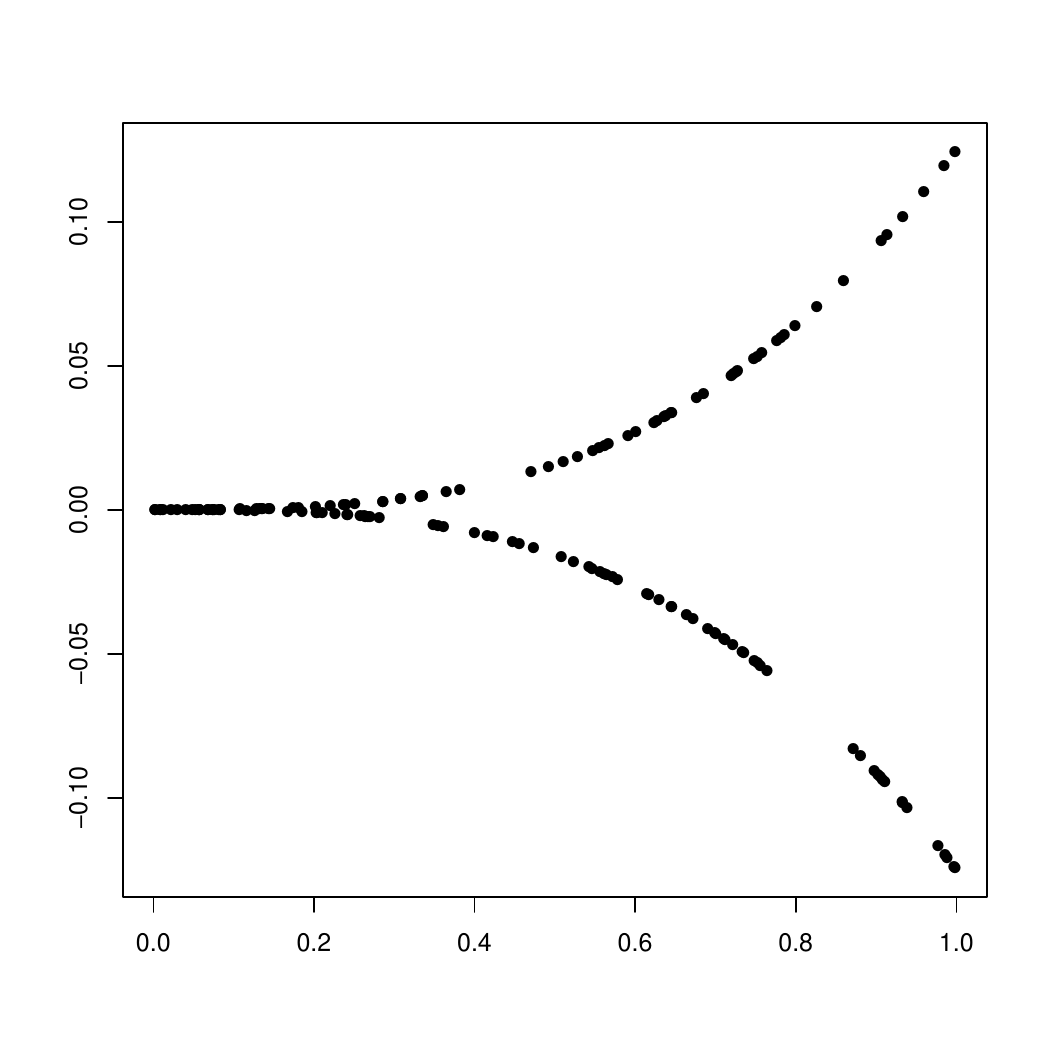}
\caption{Scatter plot of 150 couples $(x_i, y_i)$ drawn according to model \eqref{mod2}}
\label{fig:mod2}
\end{figure}

\noindent{\bf Second scenario:} we simulate, for different values of $n$, i.i.d. pairs $(X_i, Y_i)_{1 \leq i \leq n}$ according to the model
\begin{equation}\label{mod2}
 Y_i= \left(X_i \cdot 2(\varepsilon_i -0.5) \right)^3
\end{equation}
 where  $(X_i)_{1 \leq i \leq n}$ and $(\varepsilon_i)_{1 \leq i \leq n}$ are two independent sequences, the 
 $X_i$'s being  i.i.d. random variables with ${\mathcal U}([0,1])$ distribution, the $\varepsilon_i$'s being  i.i.d. random variables with a ${\mathcal B}(0.5)$ Bernoulli distribution (see Figure \ref{fig:mod2}). 
 Here, again,
 one can easily see that $\rho=\tau=\rho_S=0$. Hence $X_i$ and $Y_i$ are not correlated in the sense of Pearson, Kendall or Spearman; but of course, they are not independent.
 
We simulate $N=3000$ samples of size $n$ according to model \eqref{mod2}, and for each test we indicate the frequency of rejection of $H_0$ at level $5\%$. We consider the same  seven tests as in the first scenario. The results are given in Table \ref{tab2} (level $\alpha= 5 \%$).

\begin{table}
\center
\begin{tabular}{|c|c|c|c|c|c|c|c|c|c|c|}
\hline
 $n $  &    30  & 40 & 50 & 60 & 70 & 80  & 90 & 100 & 200 & 300 \\
\hline 
usual P  & 0.162  &  0.159 &  0.158 & 0.154 & 0.148 & 0.146 & 0.154 &0.153 & 0.141 &0.142\\
\hline
robust P &0.072 &0.064 & 0.06 &0.057 & 0.056 & 0.058 &0.058&0.05 &0.049  & 0.052\\
\hline
usual K &0.231 &0.236 & 0.236 &0.242 & 0.249 & 0.248 &0.255&0.247 &0.245  & 0.246\\
\hline
robust K &0.082 &0.074 & 0.063 &0.06 & 0.06 & 0.059 &0.06&0.059 &0.054  & 0.05\\
\hline
usual  S &0.138 &0.143 & 0.133 &0.132 & 0.142 & 0.139 &0.148&0.14 &0.142  & 0.135 \\
\hline
robust S &0.113 &0.099 & 0.085 &0.079 & 0.071 & 0.071 &0.07&0.068 &0.059  & 0.056 \\
\hline
KS indep &0.956 &1 & 1 & 1 & 1 & 1 & 1& 1 & 1  & 1 \\
\hline
\end{tabular}
\caption{Frequencies of rejections of the seven tests at level $5\%$ for model  \eqref{mod2}.}
\label{tab2}
\end{table}

From Table 2 we can make much the same comments and observations as for scenario 1 (although models \eqref{mod1} and \eqref{mod2} are quite different). The correlation tests are all poorly calibrated. Robust tests always have a lower rejection frequency than usual tests. In fact, for robust P, the frequencies of rejection are always between $5\%$ and $6\%$ for $n\geq 50$; for robust K, the frequencies of rejection are always between $5\%$ and $7\%$ for $n\geq 50$; for robust S, the frequencies of rejection are always between $5\%$ and $7\%$ for $n\geq 90$.  Our last observation is that, unsurprisingly, the independence test detects very well the non-independence of $X$ and $Y$, systematically as soon as $n\geq 40$.\\

\noindent{\bf Conclusion:} We have seen, through two really different scenarios of simulations, that if we want to test non-correlation, it is preferable to use robust tests. Indeed the usual tests can lead to a false positive rate that is much too high. One could object that the usual correlation tests can be used to test the null hypothesis $H_0$: $X$ and $Y$ are independent. But our simulation study reveal that the usual correlation tests are not consistent to test this hypothesis (their rejection frequency does not tend to 1 when $n$ tends to infinity), whereas the Kolmogorov-Smirnov type test is consistent.\\

\noindent{\bf Remark:} 
Concerning the particular case of Pearson's test of correlation: if we know that the couple $(X,Y)$ is Gaussian, then the usual test will be well calibrated, and more powerful than the robust version. Note however that, to our knowledge, there is no practical way to test whether the pair $(X,Y)$ is Gaussian (which is a much stronger assumption than assuming that $X$ and $Y$ are Gaussian).

\section{Robust test for the equality of variances}\label{Sec4}

Let $(X_1,Y_1), \ldots, (X_n,Y_n)$ 
be independent and identically distributed (i.i.d) copies of the random pair $(X,Y)$, where $X$ is real-valued random variable and $Y$ is a categorical random variable with $p$ different levels $L_1, \dots, L_p$.
Assume that ${\mathbb E}(X^2)< \infty$, and let $\mu_i= {\mathbb E}(X|Y=L_i)$ and $V_i=\text{Var}(X|Y=L_i)$. The problem of testing the hypothesis $H_0 : \mu_1= \cdots = \mu_p$ is called ANOVA (Analysis of Variance) in the literature. If we do not assume that the variances $V_i$ are all equal, this problem has been solved satisfactorily by  James  \cite{J} and Welch  \cite{We2}. The procedure proposed by James and Welch works for any sample size $n$ when the conditional distribution of $X$ knowing $Y$ is Gaussian, but also without any additional assumptions when $n$ is large enough (the test is asymptotically well calibrated).
More precisely, denoting by $JW_n$ the James-Welch test statistic, we know that :
under $H_0$, $(p-1)JW_n$ converges in distribution  as $n \rightarrow \infty$ to a $\chi^2(p-1)$ distribution.
\noindent
For the ANOVA test of James-Welch (not assuming equal variances), the corresponding R function of the {\it{stats}} package is: \texttt{oneway.test()}.\\

\noindent
Consider now the problem of testing the equality of conditional variances, that is 
$$
H_0 : V_1= \cdots = V_p  \quad \mbox{against}  \quad H_1 : \exists\, i,j  ~\mbox{such that}~ V_i \neq V_{j} 
$$ 
We will need some notations in order to proceed. Denote by $L(Y_i)$ the level of the random variable $Y_i$, by $n(L_k)$ the number of variables $Y_i$ for which $L(Y_i)=L_k$, and by
$$
\bar X_{L_k}=\frac{1}{n(L_k)} \sum_{i=1} ^{n} X_i  {\bf 1}_{Y_i=L_k}\, .
$$
\noindent
If $p=2$, to test the equality of conditional variances, one often uses the Fisher statistic
$$
 F_{n(L_1), n(L_2)}= \frac{S^2_{n(L_1)}}{S^2_{n(L_2)}}, \quad
\text{where} \quad
 S^2_{n(L_k)}=\frac{1}{n(L_k)-1} \sum_{i=1} ^{n} (X_i - \bar  X_{L_k})^2 {\bf 1}_{Y_i=L_k}\, .
$$
\indent
The interest of this statistic being that, in the case where the conditional distribution of $X$ knowing $Y$ is Gaussian, its exact conditional distribution given $(n(L_1),n(L_2))$ under $H_0$ is known: it is the Fisher distribution
$F(n(L_1)-1, n(L_2)-1)$. Fisher's test for equality of variances is constructed from the quantiles of this distribution.
But if the conditional distribution of $X$ knowing $Y$ is not Gaussian, the distribution of $F_{n(L_1), n(L_2)}$ under $H_0$ is not known.
For $p\geq 2$, one can use Bartlett's test for equality of variances \cite{Ba}, but again this test (based on the Gaussian likelihood) is not robust to non normality. We can therefore make the following conclusion: in a general context, Fisher's and Bartlett's tests are not well calibrated to test the equality of variances.
See also \cite{B} for a similar observation.\\
\indent
An alternative to Bartlett's test often mentioned in the literature is Levene's test (modified by Brown and Forsythe \cite{BF}), valid for any $p\geq 2$. This test consists of performing a classic ANOVA on the variables $|X_i-\text{Med$_n$}(X|L(Y_i))|$, where $\text{Med$_n$}(X|L_k)$ is the empirical estimator of the conditional median of $X$ given $Y=L_k$. It is quite easy to see that this procedure does not test the equality of conditional variances, but the equality of absolute deviations from the conditional medians (as it is clearly indicated in the R documentation of the package {\it{lawstat}}).\\
\indent
Assume now that the random variable $X$ has a finite moment of order 4. In order to test 
$$
H_0 : V_1= \cdots = V_p  \quad \mbox{against}  \quad H_1 : \exists\, i,j  ~\mbox{such that}~ V_i \neq V_{j}
$$
we propose to perform a James-Welch ANOVA on the variables $Z_{i}=(X_i-\bar X_{L(Y_i)})^2$. Let $T_n$ be the test statistic that we obtain by proceeding in this way. 
It is then easy to deduce from the James-Welch test statistic 
that, under $H_0$, $(p-1)T_n$ converges in distribution as $n \rightarrow \infty$ to a $\chi^2(p-1)$ distribution.\\

\noindent
The corresponding R function of the {\it{robusTest}} package is:
\texttt{vartest()}.\\

\noindent {\bf Remark:} The procedure described above still works within the framework of the "fixed design" ANOVA, i.e. when we observe $p$ independent sequences $(X_{i,1})_{1\leq i \leq n_1}, \ldots, (X_{i,p})_{1\leq i \leq n_p}$ of i.i.d. variables (in this case $V_j=\text{Var}(X_{1,j}) $), provided that all $n_i$ are large enough.\\

\noindent
In the next subsection, we will illustrate the differences between the classical test  for the equality of variance and the corrected test by a small simulation study.

\subsection{Simulation study}

\begin{figure}[htbp]
\centering
\includegraphics[width=11.5 cm, height=8 cm]{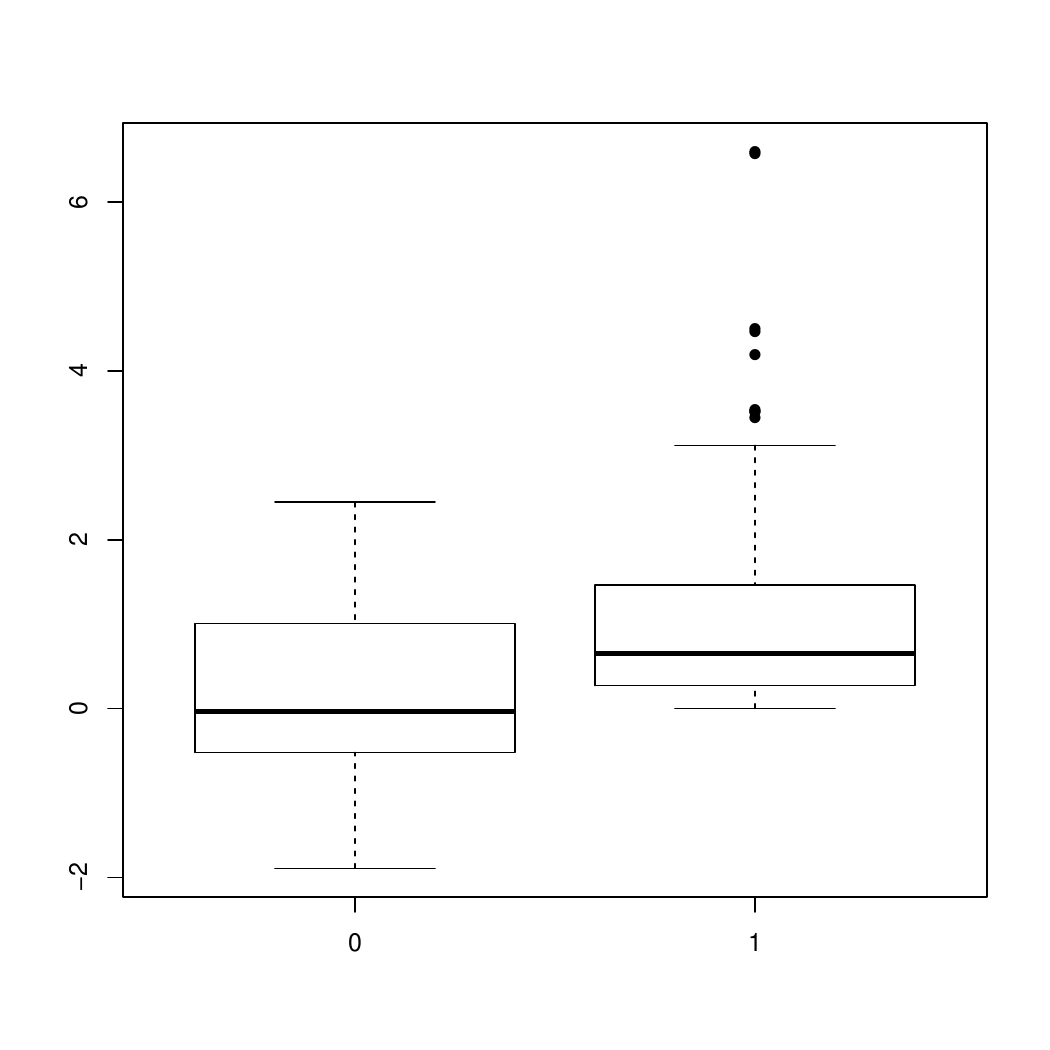}
\caption{Boxplots of the conditional distributions of $(x_i)$ given $(y_i)$ based on 180 observations 
drawn according to model \eqref{mod3}.}
\label{fig:bplot2}
\end{figure}

We simulate, for different values of $n$, i.i.d. pairs $(X_i,Y_i)_{1 \leq i \leq n}$ according to the following model:
\begin{equation}
\label{mod3}
Y_i \sim {\mathcal B}(2/3),
\end{equation}
where
the conditional distribution of $X_i$ given $Y_i=0$ is the ${\mathcal N}(0,1)$ distribution and
the conditional distribution of $X_i$ given $Y_i=1$ is the ${\chi^2}(2)/2$ distribution.
The boxplots of the conditional distributions of $(x_i)$ given $(y_i)$ based on 180 observations  drawn according to model \eqref{mod3} are shown in Figure \ref{fig:bplot2}.

Note that the conditional variances of $X$ given $Y=0$ or $Y=1$ are equal; that is, with the notations above, $V_1=V_2=1$ and the hypothesis $H_0: V_1=V_2$ is satisfied.

We simulate $N=3000$ samples of size $n$ according to model \eqref{mod3}, and for each test we indicate the frequency of rejection of $H_0: V_1=V_2$ at level $5\%$. We consider four tests : Fisher's test, Bartlett's test, Levene's test, and our test based on the James-Welch procedure (VWelch). The results are given in Table \ref{tab3} (level $\alpha= 5 \%$). 

\begin{table}[h]
\center
\begin{tabular}{|c|c|c|c|c|c|c|c|c|c|}
\hline
 $n$  &  60  & 70 & 80 & 90 & 100 & 150 & 200 & 250 & 300   \\
\hline 
Fisher &0.145  & 0.151  &  0.144 & 0.154 &  0.152 &0.153 & 0.161& 0.16& 0.161 \\
\hline
Bartlett  &0.142 &0.147 & 0.143 & 0.153 & 0.152 &0.153 &0.161 & 0.16 & 0.161 \\
\hline
Levene & 0.08 & 0.083 & 0.091 & 0.097 & 0.106 & 0.119 & 0.155 & 0.172 & 0.196 \\
\hline
 VWelch &0.057 & 0.054 & 0.056 & 0.056 & 0.055 & 0.053 & 0.056 & 0.055 & 0.053 \\
\hline
\end{tabular}
\caption{Frequency of rejection of the 4 tests at level  $5\%$ for model \eqref{mod3}}
\label{tab3}
\end{table}

From Table \ref{tab3}, our first observation is that, for this two-sample scenario, Fisher's test and Bartlett's test behave similarly, with much too high rejection frequencies hovering around $15\%$. The frequencies of rejection of Levene's test increases with $n$, from $8\%$ when $n=60$ to $19.6\%$ when $n=300$. This is not surprising, because for model \eqref{mod3}, the absolute deviations from the conditional medians are slightly different, depending on whether $Y=0$ or $Y=1$. The test VWelch, based on the James-Welch procedure is well calibrated, with frequencies of rejection between $5\%$ and $6\%$ as soon as $n\geq 60$.\\

\noindent
{\bf Conclusion:} We have seen that Fisher's and Bartlett's tests of equality of variances are not robust to non normality. Levene's test is well suited to test equality of absolute deviations from the median, but not to test the equality of variances. For testing equality of variances, it is then preferable to use the function \texttt{vartest()} of the {\it{robusTest}} package.

\section{Robust test for testing stochastic dominance}\label{Sec3}
Let $(X_1, \ldots, X_{n_1})$ and $(Y_1, \ldots, Y_{n_2})$ be two independent sequences of i.i.d. real-valued random variables (the $X_i$'s are independent copies of $X$, and the $Y_i$'s are independent copies of $Y$, the variables $X$ and $Y$ being independent). Assume moreover that the variables $X$ and $Y$ are continuous. 
We want to know if the variables $Y$ tend to take larger or smaller values than the variables $X$.
Let Med($Y-X$) be the median of $X-Y$. To answer the question, one can for example test the hypothesis
$$
   H_0: \text{Med($Y-X$)}=0 \quad \text{against} \quad H_1:  \text{Med($Y-X$)}\neq 0  \, .
$$
Mann and Whitney \cite{MW} proposed the test statistic
$$
  T_{n_1, n_2}= \frac{1}{n_1 n_2}  \sum_{i=1}^{n_1} \sum_{j=1}^{n_2} \left ( {\mathbf 1}_{X_i< Y_j} -0.5 \right )  \, .
$$
It is quite easy to see, that if $X_i$ and $Y_i$ have the same distribution, then the distribution of $T_{n_1,n_2}$ is distribution-free (i.e. does not depend on the
 common distribution of $X_i$ and $Y_i$): it suffices to reduce to two independent sequences $U_1, \ldots, U_{n_1}$ and $V_1, \ldots, V_{n_2}$ of i.i.d. random variables 
 with ${\mathcal U}([0,1])$ distributions. Consequently, if $X_i$ and $Y_i$ have the same distribution, $T_{n_1, n_2}$ follows a known and tabulated distribution.
 The Mann-Whitney test is constructed from the quantiles of this distribution. It is well known that this test is in fact equivalent to Wilcoxon rank-sum test (see~\cite{Wi}). But if $X_i$ and $Y_i$ are not identically distributed, the distribution of statistics $T_{n_1, n_2}$ has no reason to be distribution-free under $H_0$.
We can therefore make the following conclusion:
in a general context, the Mann-Whitney test is not well calibrated to test $\text{Med($Y-X$)}=0$.\\
\indent
 We can nevertheless solve (asymptotically) this problem, by considering the limiting distribution of
 $a_{n_1,n_2} T_{n_1, n_2}$ under $H_0$, for a suitable normalisation $a_{n_1, n_2}$.
 Starting from Hoeffding's decomposition of the $U$-statistic $T_{n_1, n_2}$ (see for instance \cite{vdv}, example 12.7), we see that, under $H_0$,
$$
 \frac{T_{n_1, n_2}}{ \sqrt{\frac{V_1}{n_1} + \frac{V_2}{n_2}}} \quad 
 \text{converges in distribution as $n_1, n_2 \rightarrow \infty$ to the  ${\mathcal N}(0, 1)$ distribution,}
 $$ with
$$
V_1= {\mathrm{Var}}(H_Y(X_1)) \quad \text{and} \quad V_2=  {\mathrm{Var}}(F_X(Y_1)) \, ,
$$
where $F_X(x)= {\mathbb P}(X_1 <x)$ and $H_Y(x)={\mathbb P} (Y_1 >x)$. The empirical estimators of
$V_1$ and $V_2$ are then
$$
 V_{1, n_1, n_2}=  \frac{1}{n_1-1} \sum_{k=1}^{n_1} (H_{n_2}(X_k) - \bar H_{n_2})^2  \quad 
\text{and}  \quad 
 V_{2, n_1, n_2}= \frac{1}{n_2-1}  \sum_{k=1}^{n_2} (F_{n_1}(Y_k) - \bar F_{n_1})^2
$$
where 
\begin{align*}
F_{n_1}(x) &= \frac{1}{n_1} \sum_{k=1}^{n_1}  {\mathbf 1}_{X_k < x}  & H_{n_2}(y) =\frac{1}{n_2} \sum_{k=1}^{n_2}  {\mathbf 1}_{Y_k >y} \\
\bar F_{n_1} &= \frac {1}{n_2} \sum_{k=1}^{n_2}
F_{n_1}(Y_k)  & \bar H_{n_2}  = \frac {1}{n_1} \sum_{k=1}^{n_1}  H_{n_2}(X_k)\, .
\end{align*}
Finally, under $H_0$, 
$$ 
MW_{n_1, n_2}=  \frac{T_{n_1, n_2}}{ \sqrt{\frac{V_{1, n_1,n_2}}{n_1} + \frac{V_{2, n_1, n_2}}{n_2}}} 
$$
converges in distribution
as $n_1, n_2 \rightarrow \infty$ to the  ${\mathcal N}(0, 1)$ distribution.

The rejection region of $H_0$ of the corrected Mann-Whitney test is therefore of the form
$R_{n_1, n_2, \alpha}= \{ |MW_{n_1, n_2}|> c_\alpha \}$ where $c_{\alpha}$ is the quantile of order $1-(\alpha /2)$ of the ${\mathcal N}(0,1)$ distribution, which provides a test
asymptotically well calibrated.\\

\noindent
The corresponding R function of the {\it{robusTest}} package is: \texttt{wilcoxtest(,paired=FALSE)}.

\medskip

\noindent {\bf Remark:} The procedure described above still works
when both samples are obtained from $n$ i.i.d random variables $(X_i,Y_i)_{1 \leq i \leq n}$, where the $X_i$'s are continuous random variables and the  $Y_i$'s are categorical variables with two levels $L_1,L_2$,  considering the two sub-samples of continuous variables obtained by conditioning with respect to $Y_i=L_1$ or $Y_i= L_2$.\\

\noindent
In the next subsection, we will illustrate the differences between the classical test  for testing stochastic dominance and the corrected test by a small simulation study.

\subsection{Simulations}

\begin{figure}[h]
\centering
\includegraphics[width=11.5 cm, height=8 cm]{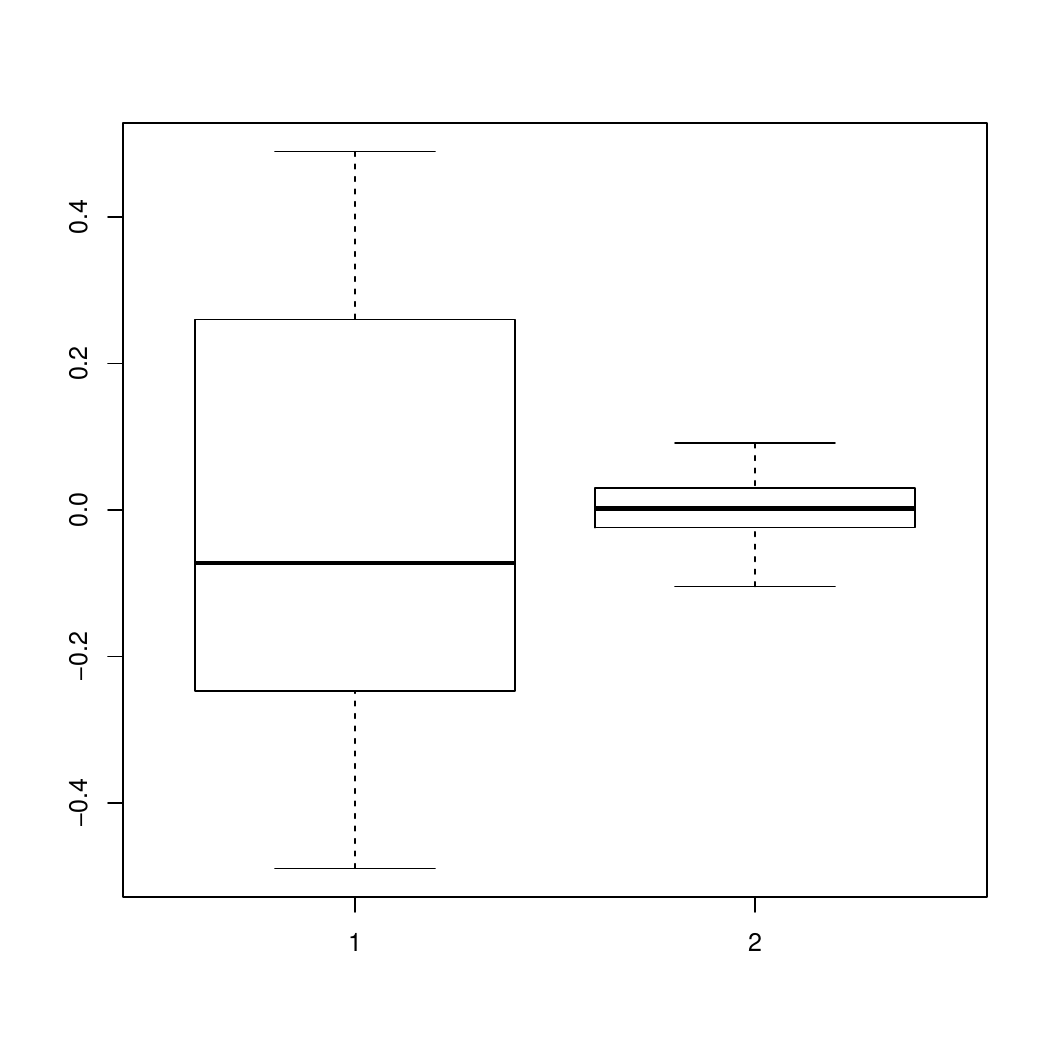}
\caption{Boxplots of 1: 60 observations drawn according to the distribution ${\mathcal U}([-0.5,0.5])$, and 2: 
180 observations drawn according to the distribution 
${\mathcal N}(0, (0.04)^2)$.}
\label{fig:bplot1}
\end{figure}

We simulate $N=3000$ samples of size $n_1$  according to the distribution ${\mathcal U}([-0.5,0.5])$,  and 3000
samples of size $n_2$  according to the distribution  ${\mathcal N}(0, (0.04)^2)$ (with standard deviation equal to 0.04), with  $n_2=3 n_1$
(see Figure \ref{fig:bplot1}).   For each test, we indicate the frequency of rejection of
$H_0$ at level $5\%$. We consider here 4 tests: the robust Mann-Whitney (robust M-W) test based on the $MW_{n_1, n_2}$ statistic, the uncorrected Mann-Whitney (M-W) test (both for testing $H_0: \text{Med($Y-X$)}=0$, which is satisfied here), Welch's test 
(which tests the equality of the expectations without assuming the equality of the variances, see the previous section), and the
two-sample Kolmogorov-Smirnov (K-S)  test  for testing the equality of distributions (see for instance \cite{Co}, pages 309-314).
The results are given in Table \ref{tab5} (level $\alpha= 5 \%$).

\begin{table}[h]
\center
\begin{tabular}{|c|c|c|c|c|c|c|c|c|c|}
\hline
 $n_1;n_2 $  
&    20;60  & 30;90 & 40;120 & 50;150 & 60;180& 70;210 & 80;240 & 90;270 & 100;300  \\
\hline 
robust M-W &0.061  & 0.064  &  0.057 & 0.058 &  0.054 &0.056 & 0.055& 0.054& 0.051 \\
\hline
M-W  &0.172 &0.171 & 0.18 & 0.166 & 0.17 &0.186 &0.168 & 0.179 & 0.161 \\
\hline
Welch & 0.057 & 0.048 & 0.051 & 0.057 & 0.49 & 0.57 & 0.053 & 0.051 & 0.051 \\
\hline
 K-S &0.998 & 1 & 1 & 1 & 1 & 1 & 1 & 1 & 1 \\
\hline
\end{tabular}
\caption{Frequency of rejection of the four tests at level $5\%$ for $n_1$ pairs drawn according to the distribution ${\mathcal U}([-0.5,0.5])$, and
  $n_2$ couples according to the distribution ${\mathcal N}(0, (0.04)^2)$.}
\label{tab5}
\end{table}

On this set of simulations (Table \ref{tab5}), we see that, for the robust M-W test and all the proposed sample sizes, the rejection frequency of $H_0$ is below
$7\%$. It is between $5\%$ and $6\%$ as soon as $ n _1\geq 40, n_2 \geq 120$. 
Welch's test is also well calibrated, with rejection frequencies all between $4.8 \%$ and $5.7\%$.
The uncorrected Mann-Whitney test is poorly calibrated, with rejection frequencies around $17\%$.
As expected, the two-sample Kolmogorov-Smirnov test detects very well the difference between the two distributions, systematically as soon as
$ n _1\geq 30, n_2 \geq 90$.\\

\noindent
{\bf Conclusion} If we want to test if Med($Y-X$)=0, it is a priori preferable to use the robust Mann-Whitney test, the usual test can lead to a false positive rate that is much too high. One could object that the usual Mann-Whitney test can be used to test the hypothesis $H_0$: $X$ and $Y$ have the same distribution. But our simulations reveal that the usual Mann-Whitney test is not consistent to test this hypothesis (its rejection frequency does not tend to 1 when $n$ tends to infinity), whereas the two-sample Kolmogorov-Smirnov type test is consistent.\\

\noindent
We also investigated how to test stochastic dominance in the case of paired samples, which will be the subject of the next section.

\section{Other tests of the {\it robusTest} package for paired two-sample}

In the {\it robusTest} package, we also provide functions to test stochastic dominance in the case of paired samples. The context is as follows: let $(X_1, Y_1), \ldots , (X_n, Y_n)$ be independent copies of the pair $(X,Y)$, where $X,Y$ are real-valued random variables. We also assume that the variables are continuous. As in the previous section, we want to know if $Y$  tends to take larger or smaller values than $X$. But here, the variables $X$ and $Y$ are a priori not independent. Classically, we consider the series of differences $D_i = Y_i-X_i$, and we are therefore reduced to a problem of univariate statistics.
We can then use the confidence interval for the median Med($D$) described in Example 21.8 of the book by van der Vaart \cite{vdv}, and based on the order statistics $(D_{(i)})_{1\leq i \leq n}$.
Let us describe this confidence interval: for $\alpha \in (0,1)$, let
$$
k_{n, \alpha}= \left [-c_\alpha \frac{\sqrt n}{2} + \frac n 2 \right ]\, ,  \quad \ell_{n, \alpha}= \left [c_\alpha \frac{\sqrt n}{2} + \frac{n}{2}\right ] \, ,
$$
the square brackets denoting the integer part, and $c_\alpha$ being  the quantile of order $1-(\alpha/2)$ of the  ${\mathcal N}(0,1)$ distribution.
Then, the interval
\begin{equation}\label{IC}
IC_{1- \alpha}= [D_{(k_{n, \alpha})}, D_{(\ell_{n, \alpha})}] \, ,
\end{equation}
is a confidence interval of Med($D$) with  asymptotic confidence level $1-\alpha$. 
As usual, a test of $H_0: \text{Med}(D)=0$ against $H_1: \text{Med}(D)\neq 0$ can be deduced from \eqref{IC}. This test is an alternative to the sign test.

For these confidence interval and test, the R function of the package {\it robusTest} is \texttt{mediantest}. This function can be used for a single sample, as well as in the case of  paired 
two-sample (in this case, the confidence interval and the test are for the quantity Med($Y-X$)). 

Note that we could have implemented  the non-asymptotic confidence interval for Med$(D)$ (which is also described in Example 21.8 of \cite{vdv}). However, some simulations suggest that, even for small samples ($25\leq n \leq 50$), the asymptotic confidence interval behaves as well as the non asymptotic one.


As another alternative to the sign test, Wilcoxon's sign and rank test is often proposed,
whose statistic is written
$$
   W_n= \sum_{i=1}^n R_i {\mathbf 1}_{D_i>0} \, ,
$$
where $R_i$ is the rank of $|D_i|$ in the sample $(|D_k|)_{1 \leq k \leq n}$.
If the variable $D$ is symmetric (which means that $D$ has the same distribution as $-D$),
Wilcoxon  \cite{Wi} showed that $|D|$ and ${\mathbf 1}_{D>0}$ are independent; we can then easily
deduce that the distribution of $W_n$ is distribution free (i.e. does not depend on the distribution of $D$). Therefore,
if  $D$ is  symmetric, the distribution of $W_n$ is known and tabulated. Wilcoxon's sign and rank test  is constructed from the quantiles of this distribution.
But if the variable $D$ is not symmetric, the distribution of $W_n$ has no reason to be distribution free under $H_0: \text{Med}(D)=0$.
We can therefore make the following conclusion:  in a general context, Wilcoxon's sign and rank test is not correctly
calibrated to test Med($D$)=0.
In fact, the sign and rank test allows to test
$$
H'_0: \text{Med($D_1+D_2$)}=0 \quad \text{against} \quad H_1: \text{Med($D_1+D_2$)}\neq 0 \, .
$$
But then again, it is not well calibrated to test this hypothesis.
We can nevertheless solve (asymptotically) this problem, by considering the limiting distribution of
 $\sqrt{n} (2U_n/(n(n-1)) -0.5)$ under $H'_0$, with
 $$
 U_n= \sum_{i=1}^n \sum_{j=1}^{i-1} {\bf 1}_{D_i+D_j >0} \, .
 $$
 Note that
 $$
 W_n= U_n + \sum_{i=1}^n {\bf 1}_{D_i>0} \, ,
 $$
 the second term on the right in the equality being asymptotically negligible with respect to $U_n$.
 Starting from the Hoeffding decomposition of the $U$-statistic $U_n$ (see for example \cite{vdv}, example 12.4),
 we see that the variable $\sqrt{n} (2U_n/(n(n-1)) -0.5)$ converges in distribution under
 $H'_0$ to the distribution ${\mathcal N}(0, V)$, where
 $$
 V=4 {\mathrm{Var}}(F(-D))
 $$
 and $F$ is the distribution function of the variables $D_i$. The empirical estimator of $V$ is therefore
 $$
   V_n= \frac{1}{n-1} \sum_{i=1}^n (F_n(-D_i)- \bar F_n)^2\,
 ~\mbox{where}~
   F_n(x)= \frac 1 n \sum_{i=1}^n {\bf 1}_{D_i \leq x} \quad \text{and} \quad \bar F_n= \frac 1 n \sum_{ i=1}^n F_n(-D_i) \, .
 $$
 Finally, under $H'_0$,
 $$
    W'_n= \frac{\sqrt{n}}{ \sqrt{V_n}} \left (\frac{2U_n}{n(n-1)} -0.5 \right ) 
 $$
 converges in distribution
when $n \rightarrow \infty$ to the distribution ${\mathcal N}(0, 1)$.

 The rejection region of $H'_0$ of the corrected sign and rank test is therefore of the form
$R_{n, \alpha}= \{ |W'_n|> c_\alpha \}$ where $c_{\alpha}$ is the quantile of order $1-(\alpha/2)$ of the distribution ${\mathcal N}(0,1)$, which provides a test asymptotically well calibrated.\\

\noindent
For the robust Wilcoxon sign and rank  test, the correspondind R function of the {\it{robusTest}} package is: \texttt{wilcoxtest(,paired=TRUE)}.

\medskip

\noindent{\bf Remark.} As we see, the Wilcoxon sign and rank test is not well suited to test $H_0: \text{Med}(D)=0$ (calibration and consistency issues) nor $H_0: $ "$D$ is symmetric" (consistency issues). If one wants to test the symmetry, one can rather use the Kolmogorov-Smirnov type statistic:
$$
K_n= \sqrt n \sup_{t \in {\mathbb R}} |F_n(t)-F_{n,-}(t)| \, ,
$$
where 
$$
F_n(t)= \frac 1 n \sum_{i=1}^n {\bf 1}_{D_i \leq t} \quad \text{and} \quad F_{n, -}(t) = \frac 1 n \sum_{i=1}^n {\bf 1}_{-D_i \leq t}\, .
$$
One can easily check that the test statistic $K_n$ is distribution-free under $H_0: $ "$D$ is symmetric", and that it
converges in distribution under $H_0$ to the supremum of a Gaussian process.

\medskip


\noindent
In the last section, we illustrate on a real data set the functions of the {\it robusTest} package to test the correlation, and we compare the outputs to those of the usual tests.

\section{Testing correlation on a real data set}

We will illustrate our results on the Evans data set which comes from  the {\it{lbref}} package (see also \cite{KK}).  These are data from a cohort study in which white males in Evans County were followed for 7 years, with coronary heart disease as the outcome of interest. The data set is a data frame with 609 rows and 9 variables. We focus on the three variables :

\medskip


CDH :
outcome variable; 1 = coronary disease



CHL :
cholesterol, mg/dl



DBP :
diastolic blood pressure, mmHg



\medskip

\begin{figure}[htbp]
\centering
\includegraphics[width=10 cm, height=8 cm]{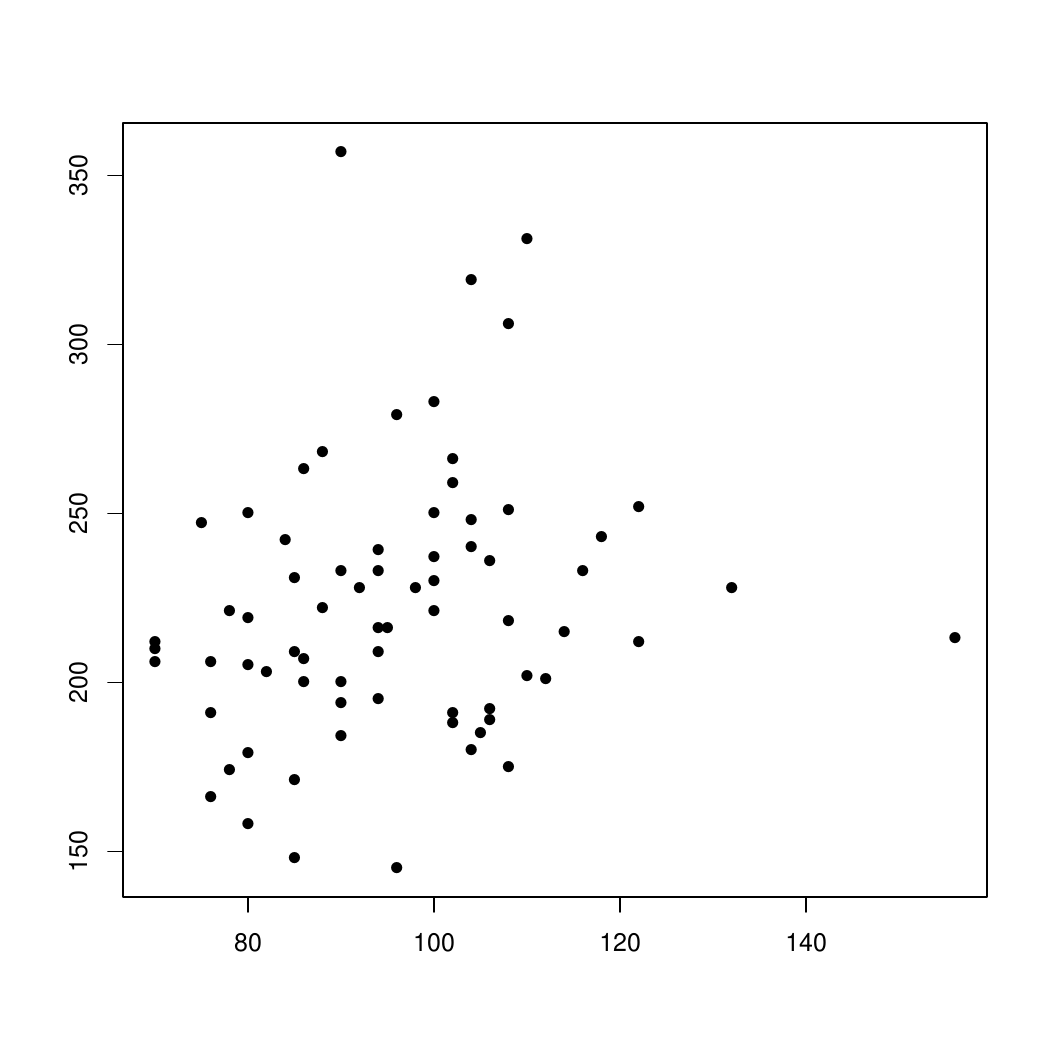}
\caption{Scatter plot of cholesterol  as a function of diastolic blood pressure in patients with coronary artery disease}
\label{fig:cloud}
\end{figure}

In this example, we restrict ourselves to the sub-sample made up of the 71 men having coronary heart disease. We propose to test the correlation between the CHL and DBP variables in men affected by this disease. The scatter plot is drawn in Figure~\ref{fig:cloud}.

\noindent
To test Pearson's correlation with the \texttt{cor.test} function of the package {\it stats},  type
\begin{verbatim}
cor.test(CHL[CDH==1],DBP[CDH==1])
\end{verbatim}
We get the output
\begin{verbatim}
Pearson's product-moment correlation

data:  CHL[CDH == 1] and DBP[CDH == 1]
t = 1.706, df = 69, p-value = 0.09251
alternative hypothesis: true correlation is not equal to 0
95 percent confidence interval:
 -0.03370795  0.41500321
sample estimates:
      cor 
0.2011779 
\end{verbatim}

\noindent
To test Pearson's correlation with the \texttt{cortest} function of the package {\it robusTest},  type
\begin{verbatim}
cortest(CHL[CDH==1],DBP[CDH==1])
\end{verbatim}
We get the output
\begin{verbatim}
Corrected Pearson correlation test

t = 2.4126, p-value = 0.0174
alternative hypothesis: true correlation is not equal to 0
95 percent asymptotic confidence interval for the correlation coefficient:
  0.0331  0.3692
sample estimates:
  cor 
0.2011779 
\end{verbatim}

We can see that the results of the two tests are very different, with a p-value of 9.25\% for the usual Pearson test, and 1.74\% for the robust Pearson test. The robust test not requiring any hypothesis on the distribution of the pair of variables, it is a priori the one that must be retained. Note that Shapiro's test on the variables CHL and DBP each time gives a p-value lower than 1\%, which implies that the observations do not come a priori from a pair of Gaussian variables. At the 5\% risk level, we conclude that there is indeed a significant correlation in the sense of Pearson.

\medskip

Let us now try to see the differences between the usual versions of the Kendall and Spearman tests and their corrected versions. Using the command \texttt{tiebreak} of the package {\it robusTest}, we see that ties are present in both series of observations. To compare the usual tests and their corrected versions, we get rid of this problem by using
\begin{verbatim}
X=tiebreak(CHL[CDH==1])
Y=tiebreak(DBP[CDH==1])
\end{verbatim}

\noindent
To test Kendall's correlation with the \texttt{cor.test} function of the package {\it stats}, type
\begin{verbatim}
cor.test(X,Y, method="kendall")
\end{verbatim}
We get the output
\begin{verbatim}
Kendall's rank correlation tau

data:  X and Y
z = 2.0003, p-value = 0.04546
alternative hypothesis: true tau is not equal to 0
sample estimates:
     tau 
0.162173 
\end{verbatim}

\noindent
To test Kendall's correlation with the \texttt{cortest} function of the package {\it robusTest},  type
\begin{verbatim}
cortest(X,Y, method="kendall")
\end{verbatim}
We get the output
\begin{verbatim}
Corrected Kendall correlation test

t = 2.4122, p-value = 0.0159
alternative hypothesis: true tau is not equal to 0
95 percent confidence interval:
  0.0304  0.2939
sample estimates:
  tau 
0.162173 
\end{verbatim}

We see a clear difference between the p-values of the two tests. That of the usual test is close to 5\%, while that of the robust test is 1.6\%. Note that these values remain globally stable when several attempts are made with the tiebreak function. The comments are exactly the same for the usual and robust Spearman tests (using \texttt{method="spearman"}): the p-value of the usual test is 4.45\% while that of the robust test is 1.72\%. Note the the p-values of the three robust tests (Pearson, Kendall, Spearman) are all below 2\%.


\bibliographystyle{plain}

\end{document}